\documentclass[journal,transmag]{IEEEtran}

\usepackage{stmaryrd}
\usepackage{cite}
\usepackage{paralist}
\usepackage{amsmath}
\usepackage{amsthm}
\usepackage{amsfonts}
\usepackage{graphicx}
\usepackage{epstopdf}
\usepackage{caption}
\usepackage{subcaption}
\usepackage{listings}
\usepackage{paralist}
\usepackage{tikz}
\usepackage[ruled,commentsnumbered,linesnumbered]{algorithm2e}
\usepackage{lipsum}
\usepackage{hyperref}
\usepackage{url}
\usepackage{multicol}
\usepackage{blindtext}
\usepackage{array}
\usepackage{floatrow}
\usepackage{lipsum}
\usepackage{wrapfig}
\usepackage{balance}
\usepackage{makecell}
\usepackage{amssymb}
\usepackage{ifthen}
\usepackage{fancyhdr}

\newcommand{\cruise}{cruise}
\newcommand{\brake}{brake}
\newcommand{\car}{\mathsf{Car}}

\newcommand{\Mathworks}{{Mathworks\textsuperscript{\textregistered}}}
\newcommand{\toolname}{{\sc DryVR}}

\newcommand{\reach}[1]{\relax\ifmmode {\sf Reach}_{ #1}\else ${\sf Reach}_{#1}$\fi}

\newcommand{\note}[1]{\ifthenelse{\boolean{comments}}{{\small \ \bf
 #1}}{}}%items for action

\theoremstyle{plain}

\newtheorem{Definition}{Definition}

\newcommand{\num}[1]{\relax\ifmmode \mathbb #1\else $\mathbb #1$\fi}
\newcommand{\nnnum}[1]{\relax\ifmmode 
	{\mathbb #1}_{\geq 0} \else ${\mathbb #1}_{\geq 0}$
	\fi}
\newcommand{\npnum}[1]{\relax\ifmmode 
	{\mathbb #1}_{\leq 0} \else ${\mathbb #1}_{\leq 0}$
	\fi}
\newcommand{\pnum}[1]{\relax\ifmmode 
	{\mathbb #1}_{> 0} \else ${\mathbb #1}_{> 0}$
	\fi}
\newcommand{\nnum}[1]{\relax\ifmmode 
	{\mathbb #1}_{< 0} \else ${\mathbb #1}_{< 0}$
	\fi}
\newcommand{\plnum}[1]{\relax\ifmmode 
	{\mathbb #1}_{+} \else ${\mathbb #1}_{+}$
	\fi}
\newcommand{\nenum}[1]{\relax\ifmmode 
	{\mathbb #1}_{-} \else ${\mathbb #1}_{-}$
	\fi}
\newcommand{\bx}{\begin{Example}}
\newcommand{\ex}{\qed\end{Example}}

\renewenvironment{IEEEbiography}[1]
  {\IEEEbiographynophoto{#1}}
  {\endIEEEbiographynophoto}

\begin{document}

\title{Road to safe autonomy with data and formal reasoning
%\thanks{The authors are supported by a grant from the National Science Foundation XXXXX.}
}

\author{
 Chuchu Fan,
 Bolun Qi, and
 Sayan Mitra
\\
 \IEEEauthorblockA{
 \{cfan10,bolunqi2,mitras\}@illinois.edu\\
 Department of Electrical and Computer Engineering,\\
University of Illinois at Urbana-Champaign.}
}

%\institute{
%\email{\{zhuang25,cfan10,mitras\}@illinois.edu}\\
%Department of Electrical and Computer Engineering,\\
%University of Illinois at Urbana-Champaign.
%\and
%\email{\{Marta.Kwiatkowska,Alexandru.Mereacre\}@cs.ox.ac.uk}\\
%Department of Computer Science, \\University of Oxford.
%}

\maketitle

\begin{abstract}
%The demand for safety assurances of safety critical cyber-physical systems has been increasing recently with the popularity of autonomous vehicles. 
We present an overview of recently developed data-driven tools for safety analysis of autonomous vehicles and advanced driver assist systems. The core algorithms combine model-based, hybrid system reachability analysis with sensitivity analysis of components with unknown or inaccessible models. We illustrate the applicability of this approach with a new case study of emergency braking systems in scenarios with two or three vehicles. This problem is representative of the most common type of rear-end crashes, which is relevant for safety analysis of automatic emergency braking (AEB) and forward collision avoidance systems. We show that our verification tool can effectively prove the safety of certain scenarios (specified by several parameters like braking profiles, initial velocities, uncertainties in position and reaction times), and also compute the severity of accidents for unsafe scenarios. Through hundreds of verification experiments, we quantified the safety envelope of the system across relevant parameters. These results show that the approach is promising for design, debugging and certification. We also show how the reachability analysis can be combined with statistical information about the parameters, to assess the risk-level of the control system, which in turn is essential, for example, for determining Automotive Safety Integrity Levels (ASIL) for the ISO26262 standard.
%   show that the verification tool can be leveraged to do risk analysis for safety standards.
%conduct rigorous safe and risk analysis. 
%Such analysis results can also be used to understand the satisfaction of systems with respect to the safety standards.  Experiments show that the reported technique is a very promising tool for safety analysis of autonomous driving systems. 
\end{abstract}


%%!TEX root = main.tex
% Introduction section
\section{Introduction}
\label{sec:intro}

%Functional safety 
There is now a race to create commercial autonomous vehicles and more advanced driving assist systems (ADAS). These safety-critical, cyber-physical systems (CPS)  use hierarchical control, long supply chains, and increasingly, machine learning algorithms. Existing design and test methodologies are inadequate for providing the needed level of safety assurances. Koopman~\cite{koopman2016challenges} argues how na\"ive test-driving for reasonable catastrophic failure rates for a fleet of vehicles can grow to hundreds of billions of miles. 
%
%
% this paper https://users.ece.cmu.edu/~koopman/pubs/koopman16_sae_autonomous_validation.pdf
%
%which is orders of magnitude higher than what is feasible. 
At the time of writing, driverless tests from Waymo and Tesla range around 100 million miles and are punctuated by  {\em disengagements}\footnote{A disengagement is an event where the human driver has to take over control from automation to prevent a hazard.}.
Precisely measuring  risks of these new technologies remains problematic, and the regulations needed for mitigating the risks are indefinite~\cite{NHTSA-plan}.
% https://www.nhtsa.gov/staticfiles/rulemaking/pdf/Automated_Vehicles_Policy.pdf

Could formal verification algorithms provide answers to these challenges?  Software model checking, for example, can find design bugs and provide rigorous safety guarantees and have proven to be practical in several domains. The computational problems related to automatically checking the safety of  CPS are notoriously difficult (undecidable). Even approximate solutions exist only for relatively simple linear models. Another fundamental problem is that traditional verification methods rely on closed-form, mathematical models of the system (e.g., differential equations and automata). In contrast, automotive systems with hundreds of modules, model-based controllers, machine learning-based units, and fine-tuned lookup tables look less like a model. Could verification algorithms work for systems with possibly incomplete models?

In this article, we report on recent research on data-driven verification that answers this question in the affirmative and we illustrate the promise of this approach with a detailed case study. The key idea is to combine traditional verification of known parts of the model, with {\em sensitivity-analysis\/} of the unknown parts that are only executable. Sensitivity analysis gives bounds on how much the states or outputs of a module change, with small changes in the input parameters. In a sequence of papers, we have presented sensitivity analysis algorithms for systems that use available knowledge of models and execution data. Based on these algorithms,  we have developed data-driven verification tools C2E2~\cite{duggirala2015c2e2} and {\toolname}~\cite{Fan2017DRYVRDV}. These tools can certify the safety of an automotive control system under a set of scenarios, or automatically find unsafe scenarios (counter-examples) that violate the safety requirements.
In Section~\ref{sec:models}, we present an overview of the inner workings of these techniques.
Previous case studies have established the effectiveness of these algorithms in verification of a benchmark engine control system,   a NASA-developed collision alerting protocol, and satellite controllers. More recently, DryVR has been used to analyze a range of autonomous and ADAS-based maneuvers~\cite{Fan2017DRYVRDV}.

We present a new detailed case study on emergency braking systems in Section~\ref{sec:scenarios}. More than 25\% of all reported accidents are rear-end crashes~\cite{kodaka2003rear}, of which around 85\% happen on straight roads. 
Emergency braking and forward collision warning  systems are becoming standard ADAS features. However, testing safety of such systems can be nontrivial~\cite{simonepresent}. The safe braking profiles for a sequence of cars on the highway depend on several factors---initial separation, velocities,  vehicle dynamics,  reaction times, road surface etc.
Our data-driven verification tool works with black-box or unknown vehicle models, using which we can prove, for example, that a given braking profile is safe for a set of scenarios characterized by the ranges of initial separation ($d$) and reaction times ($r$).   For unsafe scenarios, the tool can compute the worst case relative velocity of the collision, which determines the severity of the accident. This type of analysis can be used as a design tool for tuning the braking profiles for different highway speeds, road conditions, etc. We analyzed hundreds of scenarios to generate a safety surface that can aid such design and analysis. 
Finally, we show how data-driven verification can be used for risk analysis. ISO26262~\cite{ISO26262} classifies different control subsystems to risk levels (Automotive Safety Integrity Levels or ASILs) and prescribes process-based requirements to reduce those risks to acceptable levels. 
The risk here is broadly defined as $\mathit{severity \ of\ accident\ \times\ probability\ of\ occurrence}$.
Assessing these quantities for complex control systems, however, remains more of an art. In~\cite{simonepresent} the authors propose a method based on extensive simulations. In contrast, verification gives a provable bound on the severity of accidents for each range of $d$ and $r$ values. We show that this can be combined with statistical information about the distributions of $d$ and $r$ to obtain the risk associated with the system for a given speed and braking profile. 

In summary, we present an argument for  data-driven formal verification as a foundation for building design automation tools for safe autonomous vehicles and ADAS systems. Specifically, our algorithms combining hybrid system verification and automated sensitivity analysis of black-box models, show promise for both online monitoring and practical design-time and analysis.

%%!TEX root = main.tex

\section{Verification from Models and Data}
\label{sec:models}
Correctness of verification ultimately relies on the underlying system model which may or may not be completely known. We begin by considering dynamical systems---a simple yet very powerful modeling formalism. W have generalized it to more expressive formalisms like switched and hybrid systems, and we refer the interested readers to the article~\cite{duggirala2015c2e2}.

A {\em dynamical system\/} is represented by an ordinary differential equation (ODE):
\begin{align}
\label{eq:nsys}
\dot{x}(t) = f(x(t),u(t))
\end{align}
which describes the time-derivative, and hence, the time evolution of a vector $x$ of real-valued state variables (e.g., velocity, torque, steering angles, fuel-flow rates, etc.) with an input signal $u(t)$. 
Let us fix an input, and denote by $\xi(x_0,t)$ the {\em solution\/} of (\ref{eq:nsys}) from a particular initial state $x_0$  at time $t \geq 0$. 
The exact state is usually hard to estimate; let $K$ be  the set of possible initial states, $T > 0$ be the time horizon of interest, and $U$ be  a subset of unsafe states. $U$ are the states where speed limits are violated, emissions are excessive, collisions occur, or other bad things happen. The verification question can be written as:
\begin{align}
\label{eq:reach}
\{ x \ | \ \exists x_0 \in K,\ t \in [0, T], \xi(x_0, t) = x \} \ \cap\ U = \emptyset \ ?
\end{align}
That is, is there a behaviors of the system from $K$ that enter a bad state within $T$ time. 
The set on the left hand side is called  the {\em reachset\/}.

If $f$ is a nonlinear function or is unknown, then computing reachsets can be notoriously difficult. In fact, even if $f$ is known, $\xi(x_0,t)$ may not be computable as a closed form function of $x_0$ and $t$. However, it is  usually possible to execute or numerically simulate~(\ref{eq:nsys}), and generate data for $\xi(x_0,0), \xi(x_0,\tau), \xi(x_0,2\tau), \ldots, \xi(x_0,T)$~\cite{vnode2006}.
Data-driven verification algorithms approximate reachsets  using this simulation data and sensitivity of $\xi(x_0,t)$  to the changes in $x_0$.

 \subsection{Safety Analysis of Automatic Emergency Braking Systems}
\label{sec:scenarios} 
 An automotive control system model typically consists of several {\em modes\/}---for example, cruising, braking, shifting, merging, etc., and the software controller switches between these modes based on sensors and drivers' inputs. This gives rise to a  {\em hybrid\/} system that combines ODEs with an automaton that defines the allowed mode switches. 
%\label{sec:scenarios}
%\sayan{introduce the case study in detail. General setup, the figure, what is beign abstracted out. 
%	The state variables, the input: acceleration profile. The relevant initial conditions, the time horizon, and the unsafe set. Cite Simone's document, connect with ISO.}

\begin{figure}[tbhp!]
    \centering
 %   \begin{subfigure}[b]{0.56\textwidth}
 %       \includegraphics[width=\textwidth]{figs/two_cars}
 %       \caption{Two cars following scenario}
  %      \label{fig:two_car_following}
  %  \end{subfigure}
    %add desired spacing between images, e. g. ~, \quad, \qquad, \hfill etc. 
      %(or a blank line to force the subfigure onto a new line)
  %    \vspace{5pt}
%    \begin{subfigure}[b]{0.9\textwidth}
        \includegraphics[width=\textwidth]{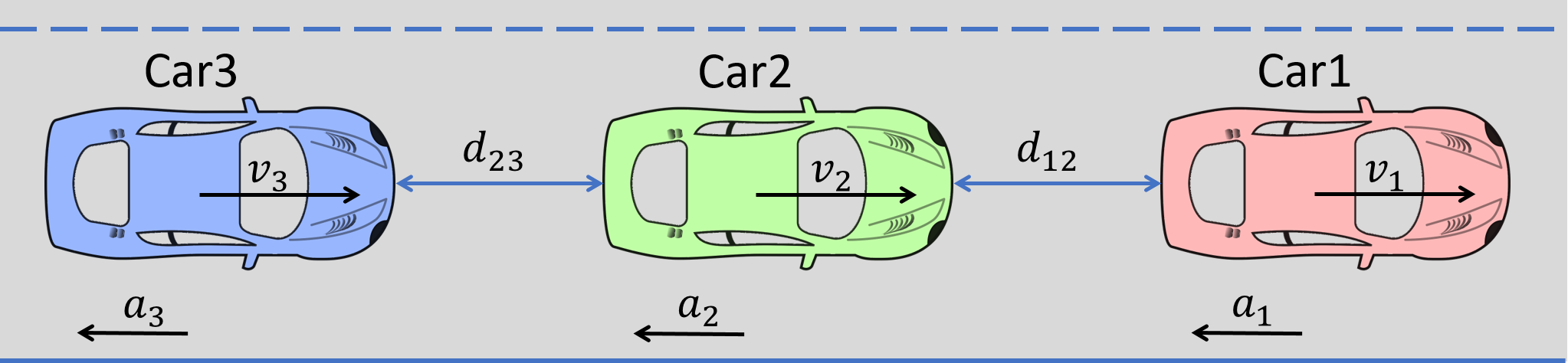}
%        \caption{Cars cruising and braking in a single lane configuration.}
%        \label{fig:three_car_following}
%    \end{subfigure}
    \caption{Cars cruising and braking in a single lane configuration.}\label{fig:reachtubes}
    \label{fig:aeb}
\end{figure}

Consider for example an Automatic Emergency Braking (AEB) system (Figure~\ref{fig:aeb}): 
$\car 1, \car2$ and $\car 3$ are cruising down the highway with zero relative velocity and certain initial relative separation;  $\car 1$ suddenly switches to a braking mode and starts slowing down according to a  certain deceleration profile. Irrespective of whether $\car 2$ is human-driven, AEB-equipped, or fully autonomous, certain amount of time elapses,  before $\car 2$ switches to a braking mode. We call this the {\em reaction time} $r$. Similarly, the mode switch for $\car 3$ happens after a delay. Obviously, $\car 2$'s safety is in jeopardy: if it  brakes ``too hard'' it will be  rear-ended and if it is ``too gentle'' then it would have a forward collision. The envelope of  safe (no collision) behaviors depend on all the parameters: initial separations, velocities, braking profiles, and reaction times. 
It is easy to see that if we can solve the safety verification problem described above, then we can also  compute this envelope and  determine whether a given AEB system is safe over a range of scenarios.

Delving deeper into this system, we remark that the exact vehicle dynamics and braking profiles are typically complex or unknown, but we can simulate the system. For the results in Sections~\ref{sec:asil},\ref{sec:int}, for example, we used a black-box vehicle model from {\Mathworks}  demo~\cite{carmodel}. 
Each car can be viewed a hybrid system: it has several continuous state variables: deceleration rate $a(t)$, velocity $v(t)$ and position $s(t)$ and two modes: $\cruise$ and $\brake$. In the $\cruise$ mode, the car maintains a constant speed, and in the $\brake$ mode, it decelerates according to a certain braking profile for $a(t)$, which is an input to the system. 
The initial set $K$ of the system is defined by the uncertainties in the initial vehicle velocities ($v_1(0), v_2(0), v_3(0)$) and the initial separations  $d_{12}, d_{23}$ between the vehicles. 
The unsafe set $U$ corresponds to states where there is a collision, that is, the separation between a pair of cars is less than some threshold. In case of collisions, we would be interested in the maximum possible relative velocity just before the collision, which strongly influences the severity of the accident.

%Therefore, the unsafe set $U$ is $\exists t \geq 0, |s_1(t) - s_2(t)| < \theta$.

The key parameters we will consider in this paper are 
the reaction time or the delay $r$ in switching to the braking mode, 
the initial separation $d$ between the cars.

%to avoid the potential collision, injuries of passengers and any other damage to the vehicle. The input to the cars are the deceleration profiles.

%\subsection{Verification of car-following scenarios}

\section{Sensitivity Analysis and Verification Algorithms}

Data-driven  verification algorithms  rely on computing reachsets from models and simulation data. First we discuss a situation where a lot of information about the system model is available, and this is exploited by the verification algorithm. Then step by step we will drop these assumptions and arrive at algorithms that only use the system as a black-box. 
%s from the initial set $K$, with an adequate degree of precision. The reachtubes are defined as a sequence of tim-stamped sets $(R_0, 0),(R_1,t_1),\dots,(R_n,t_n)$ satisfying
%\begin{inparaenum}[(1)]
%	\item Each $R_i \subseteq \reals^n$ is a compact subset of $\reals^n$.
%	\item The last time $t_n = T$ and for each $i$, $0 \leq t_i - t_{i-1} \leq \tau$.
%	\item For any $x_0 \in D$, and any time $t \in [t_{i-1}, t_i]$, the solution $\xi(x_0,t) \in R_i$.
%\end{inparaenum}
%

\subsection{Verification Algorithm}
\label{sec:algo}
As mentioned in the introduction, the key idea is to use simulations to determine the sensitivity of the system's solution $\xi(x_0,t)$ to changes in initial conditions $x_0$. 
%Using sensitivity analysis to compute the reachtube from finitely many simulation traces has been popular recently as a systematic way for state space exploration~\cite{duggirala2015c2e2,fan2016automatic,Fan:ATVA2015,chuchuEMSOFT}. 
%
%The authors introduce the {\em sensitivity matrix \/} with respect to initial state $x_0$ at time $t$ as $s_{x_0} \triangleq \frac{\partial \xi(x_0,t)}{\partial x_0}$.  It is shown that $\dot s_{x_0}(t) = J_f(x_0,t)s_{x_0}(t)$, where $J_f$ is the Jacobian matrix of $f$. Then $\|s_{x_0}(t)\|\delta$ is used to bound the distance $\|\xi(x,t)-\xi(x_0,t)\|$ for $x \in B_{\delta}(x_0)$ at time $t$. It is shown that this upper-bound holds for linear time varying systems. 
%For general nonlinear systems, $\|s_{x_0}(t)\|\delta$ has a quadratic error term with respect to $\delta$, and therefore, the above technique does not provide any formal guarantees for nonlinear systems in general. In \cite{dang2008sensitive} this technique is extended to nonlinear systems subject to disturbances as inputs and uncertainty in the initial conditions to obtain an approximation that ignores the higher order terms. 
%
%We use the following 
Formally, sensitivity is quantified by the the following notion discrepancy.  
\begin{Definition}
	\label{def:disc}
	A uniformly continuous nonnegative function $\beta$ is a {\em discrepancy function\/} of~(\ref{eq:nsys})  if
	\begin{inparaenum}[(a)]
		\item for any pair of states $x, x'$, and any time $t >0$,
		\begin{eqnarray}
		\|\xi(x,t) - \xi(x',t)\| \leq \beta(x,x',t), \mbox{and}
		\label{eq:df1}
		\end{eqnarray}
		\item for any $t$, as $x \rightarrow x'$, $\beta(\cdot,\cdot,t) \rightarrow 0$.
	\end{inparaenum}
\end{Definition}

Recall, the safety verification problem of Equation~(\ref{eq:reach}).
Assuming that a discrepancy function $\beta$ is available for the system~(\ref{eq:nsys}), the safety verification algorithm for~(\ref{eq:nsys}) proceeds as follows:
%\sayan{Needs cleaning up.}
\begin{enumerate}
	\item Compute a $\delta$-cover $C = \{x_i\}_{i=1}^k$ of the initial set $K$, i.e.,  $K \subseteq \cup_i B_\delta(x_i)$.
	\item For each $x_i \in C$, a  simulation $\psi(x_i)$ from $x_i$ is computed. 
	\item The simulation $\psi(x_i)$ from $x_i$ is expanded by a factor given by the discrepancy, such that this expanded simulation $R(\psi(x_0),\beta,\delta)$ over-approximates all the states reachable from a  $\delta$-neighborhood of $x_i$ over the time horizon $T$.
	% $\beta(x_i,..)$, that is,  let
	%\[
	%\beta_{\mathit max,i,j} = \max_{x \in B_\delta(x_i), t \in [t_j,t_{j+1}]} \beta(x_i,x,t).
	%\] 
	%The reachtube from $B_\delta(x_i)$  for the interval $[t_j, t_{j+1}]$ is then computed as $\mathit{hull}(R_j, %R_{j+1}) \oplus \beta_{\mathit max,i,j}$.
	\item If this reach set $R(\psi(x_0),\beta)$ is disjoint from the unsafe set $U$ then $x_i$ is removed from the cover $C$.
	Else if any interval of the  simulation $\psi(x_0)$ is contained in $U$  then output {\bf Unsafe\/} and $\psi(x_0)$ serves as a counter-example or bug trace. 
	Otherwise, if neither case holds, then $x_i$ is replaced in $C$ by a finer cover of the $\delta$-neighborhood of $x_i$.
	\item If the cover $C$ becomes empty, then output {\bf Safe\/}. 
\end{enumerate}

\begin{figure}[tbhp!]
	\centering
	\includegraphics[width=0.35\textwidth]{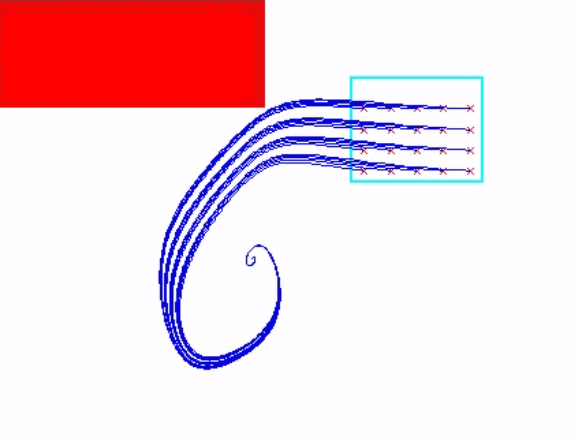}
	\includegraphics[width=0.35\textwidth]{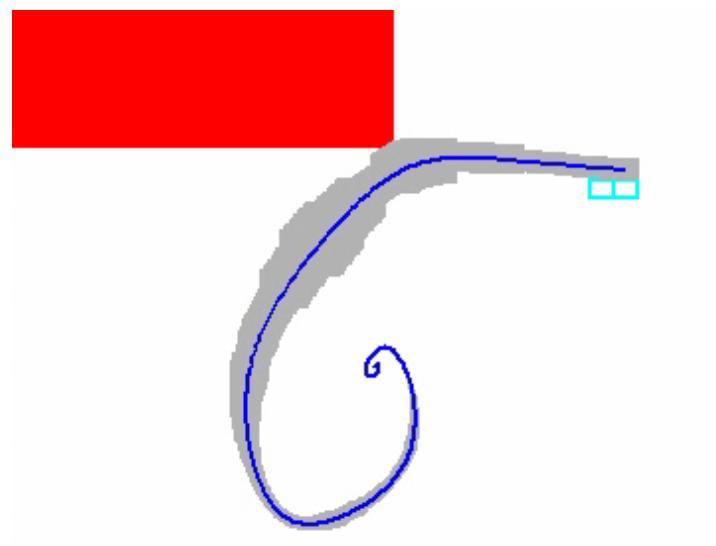}
	
	\includegraphics[width=0.35\textwidth]{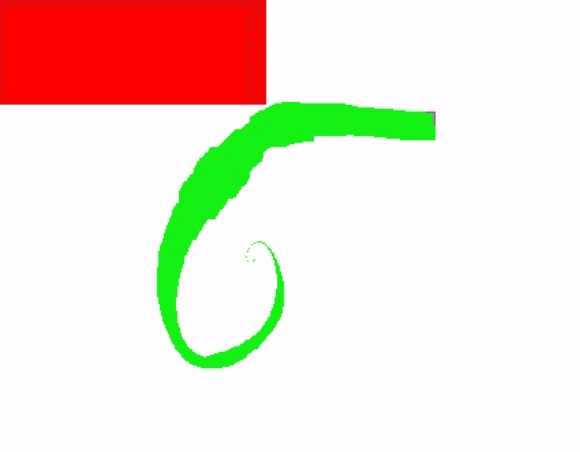}
	\includegraphics[width=0.35\textwidth]{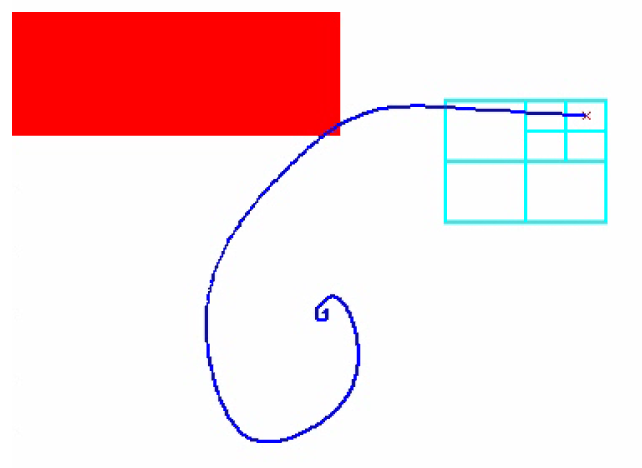}
	\caption{\small Conceptual demonstration of data-driven verification. Red rectangle: Unsafe states, Cyan rectangle: Uncertainty space. Simulations (blue lines) cannot  guarantee safety, but simulation data together with sensitivity analysis  can give coverage (grey region) to prove safety (green region) or help identify bugs quickly.}
	\label{fig:c2e2}
\end{figure}

This simple algorithm computes increasingly finer covers of $K$ until the reachsets $R(\psi(x_0),\beta,\delta)$ from each of the elements in the cover are inferred to be disjoint from the unsafe set $U$ or a counter-example simulation is discovered. The second property of the discrepancy function ensures that as the elements in the cover become finer, the over-approximation of the computed reachsets becomes more precise. 
A simple 2-D illustration of the  verification algorithm in action is shown in Figure~\ref{fig:c2e2}.
%If $\beta$ meets the two conditions for any pair of states $x,x'$ in a compact subset $K\subsetneq \reals^n$ then it is called a {\em $K$-local discrepancy function.} 
%

Essentially the same idea can be made to work for switched and hybrid models like the emergency braking scenarios described in Section~\ref{sec:disc} (see~\cite{duggirala2015c2e2} for details). The main complication is that because of the over-approximations in the computed reachsets, we have to keep track of spurious mode changes. 
% the  algorithm has to carefully track both real reachset---for producing counter-examples, and  also use  all candidate reachset for quickly proving safety. 

The  algorithm has been proved to be sound and relatively complete in~\cite{duggirala2015c2e2}. Soundness means that whenever the Algorithm returns {\bf Safe\/} or {\em Unsafe\/}, then the system is indeed safe or unsafe, respectively. For example, if the algorithm returns safe for a AEB scenario with certain  braking profiles, then there is guaranteed to be no collisions until all the cars stop. If the algorithm returns {\bf Unsafe\/}, then there are particular parameter values, such as the initial separations and reaction times, that leads to a collision. In this case, our tool will compute an upper-bound on the maximum relative velocity of all collisions that can arise from the considered range of parameter values.
Relative completeness means the algorithm will always terminate whenever the system is either robustly safe or robustly unsafe.

%
%In~\cite{althoff2008reachability} the authors present a different approach based on linearizing the nonlinear system locally, and bounding the linearization error by Lagrange remainders. The  definition of discrepancy function can be seen as a generalization of the incremental stability~\cite{angeli2002lyapunov}. The incremental Lyapunov  function  can be used as a discrepancy function when a system is incrementally stable. 
%
%The related notion of contraction~\cite{lohmiller1998contraction} is defined as the region in which the eigenvalues of the symmetric part of the Jacobian is uniformly negative. Contraction metrics introduced in \cite{lohmiller1998contraction} is  used~\cite{DMVemsoft2013} to perform sound and relative complete bounded invariant verification of  nonlinear systems.

\subsection{Computing Discrepancy}
\label{sec:disc}
The key ingredient for the above verification algorithm is a good discrepancy funciton $\beta$. 
One can get an exponentially growing discrepancy function $\beta(x,x',t)= ||x - x'||e^{Lt}$, where  $L$ is a Lipshitz constant for $f$, but even for moderately large $L$,  the reachset  over-approximations  $R(\psi(x_0),\beta,\delta)$  blow-up, and the  verification algorithm would get clogged by large number of refinements. 

Methods for computing tight discrepancy functions for linear ODEs were presented in~\cite{duggirala2015c2e2,donze2010breach}, but the problem remained open for general nonlinear models. 
In~\cite{chuchuEMSOFT}, an approach was presented for nonlinear systems that aimed to compute local discrepancy functions that are relevant only over small parts of the state space. It was shown that this preserves the soundness and relative completeness of the verification algorithm, and this approach lies at the core of our first data-driven verification tool for C2E2~\cite{duggirala2015c2e2}.

Previous methods for computing discrepancy in \cite{chuchuEMSOFT,donze2010breach} rely on availability of a closed-form system model (i.e. the dynamical function $f$). In many practical control systems, the model is at least partly unavailable or it is too complicated for deriving a closed-form description.  
%\mitras{make this a new section after section C. Describe the white-box black-box in a couple of sentences.}
In this case, hybrid control systems can be described by combining a black-box simulator for trajectories and a white-box transition graph specifying mode switches. When we only have access to a black-box simulator, a probabilistic algorithm  can be used to learn the parameters of exponential discrepancy functions from simulation data. This is the basis for our new data-driven verification tool DryVR~\cite{Fan2017DRYVRDV}.

The DryVR algorithm transforms the problem of learning the parameters of the discrepancy function to the problem of learning a linear separator for a set of points in 2-dimensions that are obtained from transforming the simulation data. The idea of the algorithm works as follows:
\begin{inparaenum}[(1)]
\item Draw $m$ samples of initial states $x_i,i=1,\dots,m$ from initial set $K$,
\item Simulate the black-box simulator to get sampled traces $\xi(x_i,t_k), i = 1,\dots,m, k = 0,\dots, N$, where $t_N = T$ is the time bound,
\item Minimize $\int_{t = 0}^{T} c e^{\gamma t}$ such that $c e^{\gamma t}$ with $c,\gamma$ a scalar value is a valid discrepancy function for any pair of traces $\xi(x_i,t),\xi(x_j,t), i,j = 1,\dots,m$ and for any time $t = t_k, k = 0,\dots, N$. 
\end{inparaenum}
With the PAC-learnability of concepts with low VC-dimension, it can be shown that for $\delta,\epsilon>0$, if $m \geq \frac{1}{\epsilon} \ln \frac{1}{\delta}$, then with probability $\leq \delta$, the constructed discrepancy function $c e^{\gamma t}$ does not work for more than $\epsilon$ fraction of the points in the initial set $K$. 
Assuming that the discrepancy function is correct, the DryVR verification algorithm gives the same soundness and relative completeness quarantees as stated in Section~\ref{sec:algo}. Our experiments suggest that a few dozen simulation traces are adequate for learning discrepancy functions with nearly 100\% correctness, for typical automotive models.

\subsection{Determining Severity of Collisions using Reachability}
\label{sec:reachtubeanalysis}
Let us consider the AEB systems discussed in Section \ref{sec:scenarios} and see how reachability analysis can guarantee safety of scenarios and compute worst-case  collision velocities. 

%Let the uncertainty of the initial separation between the vehicles  be the range $[d_{l},d_{u}]$ and the reaction time of the follower car $\car 2$ to be in the range $[r_{l},r_{u}]$. 
%%We set the initial position of $\car 2$ $s_2(0)$ to be $0$ and the initial position of $\car 1$ $s_1(0)$ to be in the range $[d_{l},d_{u}]$. 
%That is, $\car 1$ will transit from cruise mode to the brake mode at certain time $t_{b}$, then $\car 2$ will transit from $\cruise$ to $\brake$  between time $[t_{b} + r_{l}, t_{b}+r_{u}]$. 
%
%Fix the ,

%Therefore, the transition to the entire system is 

 \begin{figure}[tbhp!]
    \centering
    \begin{subfigure}[t]{0.46\textwidth}
        \includegraphics[width=\textwidth]{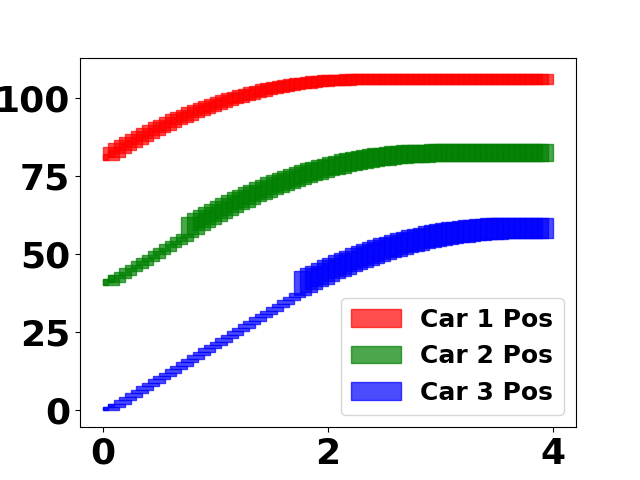}
        \caption{Safe case: reachtubes of position are separated by a distance $\geq 2$ for any time.}
        \label{fig:reachtube_safe}
    \end{subfigure}
    ~~
    %add desired spacing between images, e. g. ~, \quad, \qquad, \hfill etc. 
      %(or a blank line to force the subfigure onto a new line)
    \begin{subfigure}[t]{0.46\textwidth}
        \includegraphics[width=\textwidth]{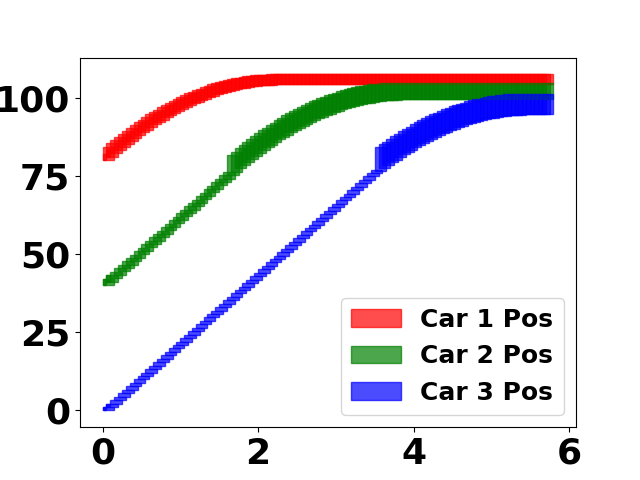}
        \caption{Unafe case: at least a pair of reachtubes of position for some time are contained in the unsafe region ($\leq 2$ separations).}
        \label{fig:reachtube_unsafe}
    \end{subfigure}
    \caption{Safety of the AEB system. Horizontal axis is time in seconds, vertical axis is position in meters.}\label{fig:reachtubes}
\end{figure}

  %{\toolname} automatically verifies  the safety of this system in $3$ to $50$ seconds on a standard laptop.
 Fix the initial velocities and braking profiles for all the cars, fix the ranges for initial separations and reaction times,
  if {\toolname} returns safe (and the learned discrepancy is correct), then it means the distance between any two cars is always larger than the threshold value $\theta=2$ meters 
  % give a concrete value in meters.
  at all times, before the cars stop. Figure~\ref{fig:reachtube_safe} shows the projection of the reachsets plotted against time for the entire range of initial conditions; the range of positions  for $\car 1$ (red),$\car 2$ (green) and $\car 3$(blue ) are separated by at least  $2$ meters. If  {\toolname} returns unsafe, then it also computes  parameter values (initial separations $d_{12},d_{23}$, reaction times $r_2,r_3$) that lead to a state where the cars have less than  $2$ meters separation. In Figure~\ref{fig:reachtube_unsafe}, the reachsets for position overlap indicating a collision and in this case the tool also over-approximates the worst case relative velocity in the collision. For example, in the particular example the worst case collision velocity between $\car 1$ and $\car 2$ is $9.0(m/s)$.
  % shown in the Figure \ref{fig:two_cars_hm}, the non-green regions are unsafe scenarios.
  %, and the maximum possible collision velocities are computed.

%\section{Applications in safety verification}
%\label{sec:exp}

\section{Risk analysis for ASIL}
\label{sec:asil}
Reachability analysis can be used for determining risk levels of an automotive control system. 
According to ISO 26262 ASIL classification, risk is broadly defined as $\mathit{severity\ of\ accident} \times \mathit{probability\ of\ occurence}$. For the AEB system with 2 cars, the severity is largely determined by the relative velocity of collision, which is approximated from the above reachability analysis. 

\begin{figure}[tbhp!]
    \centering
    \begin{subfigure}[b]{0.48\textwidth}
        \includegraphics[width=\textwidth]{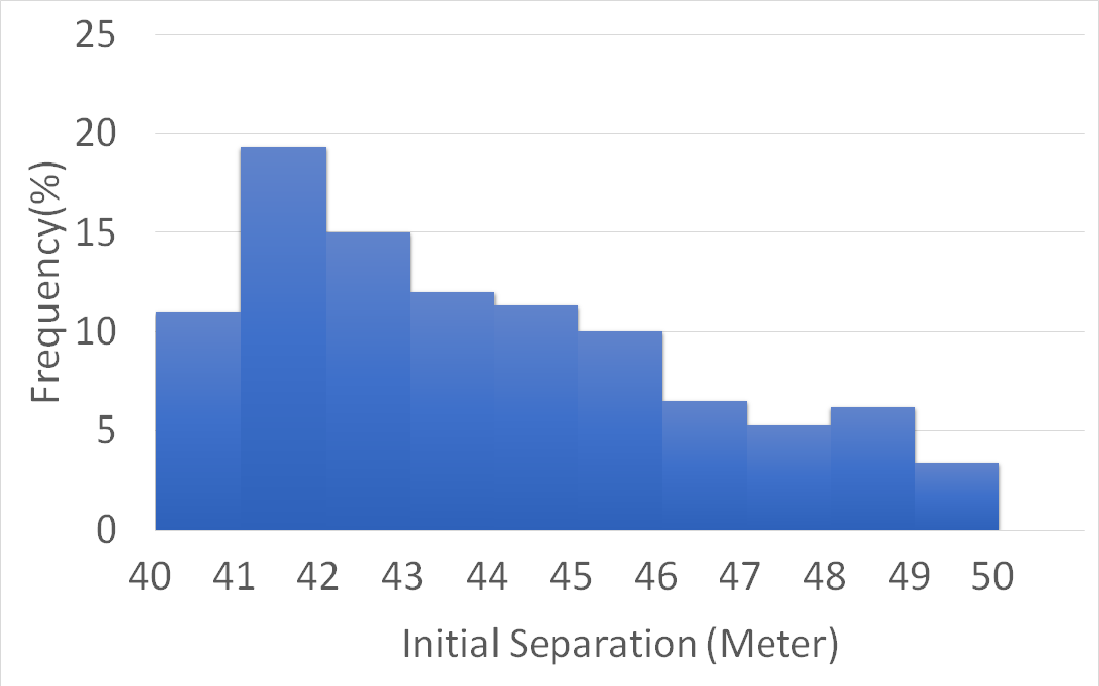}
        \caption{Initial separation distribution}
        \label{fig:dist_distri}
    \end{subfigure}
    ~
    %add desired spacing between images, e. g. ~, \quad, \qquad, \hfill etc. 
      %(or a blank line to force the subfigure onto a new line)
    \begin{subfigure}[b]{0.48\textwidth}
        \includegraphics[width=\textwidth]{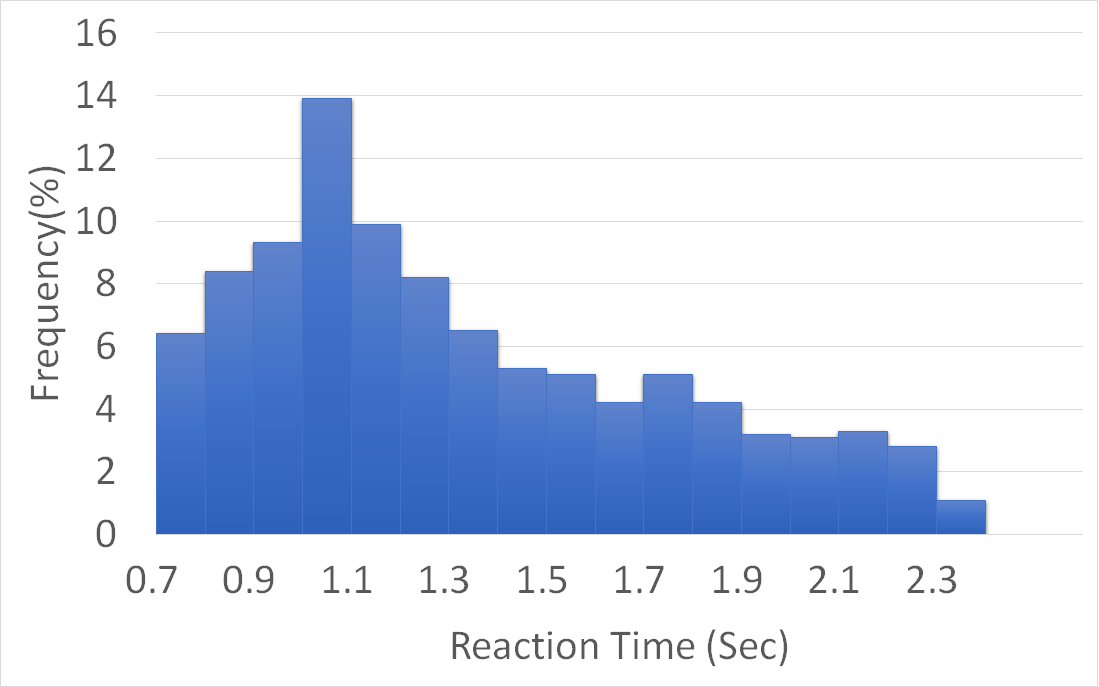}
        \caption{Reaction time distribution}
        \label{fig:react_distri}
    \end{subfigure}
        \begin{subfigure}[b]{0.99\textwidth}
        \includegraphics[width=\textwidth]{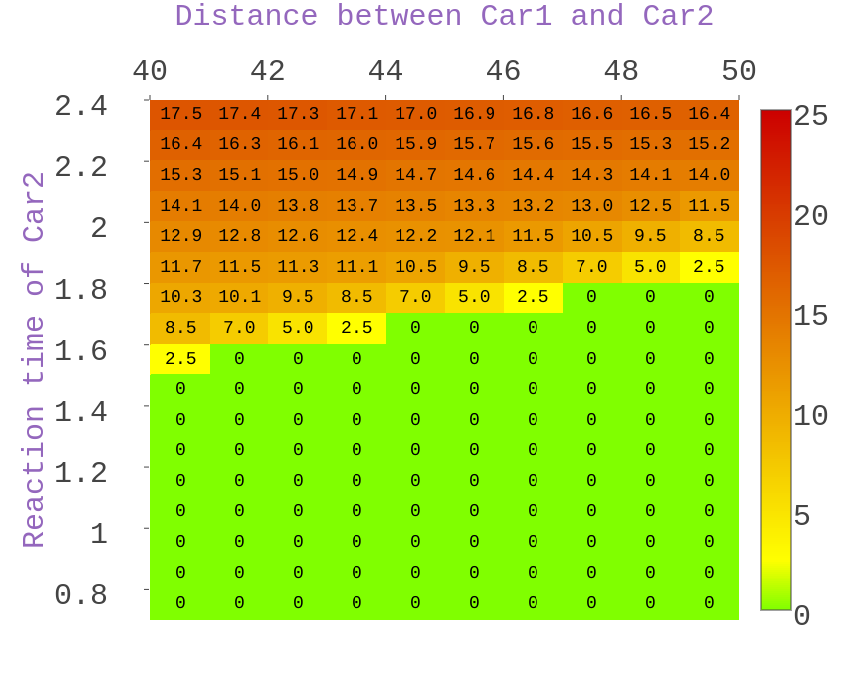}
        \caption{Worst case relative velocities (m/s) for the collisions. Braking profiles are fixed ($\car 1$: mild brake, $\car 2$: medium brake) and initial velocities are $30(m/s)$.}
        \label{fig:wcrvc}
    \end{subfigure}
    
    \caption{AEB of two cars: probability and severity}\label{fig:distributions}
\end{figure}

The probability of occurrence depends on the probability distributions on the parameters ($d, r,$ etc.). In general, these distributions can be complicated. As a starting point, the preliminary study presented in~\cite{simonepresent}, use empirical observations to construct distributions on initial separation ($d$) which turns out to be a skewed Gaussian with the mean dependent on the car speed. Similarly, the reaction time distribution is also a skewed Gaussian. Examples of such  distributions are built using~\cite{simonepresent,carparameters} and shown in Figures~\ref{fig:dist_distri} and~\ref{fig:react_distri}, where the separation $d$ ranges over  $[40,50]$ meters and reaction time $r$  ranges over  $[0.7, 2.4]$ seconds. 

%Instead of arguing actual skewed Gaussian distributions \cite{carparameters} of $d$ and $r$ for different settings, we will mainly focus on analyzing the severity of collision for given ranges of $d$ and $r$. 
%Assume that some distribution of $d$ is shown in Figure \ref{fig:dist_distri} and value of interest lies in the interval $[40,50]$ meters. Similarly, assume some distribution of $r$ is shown in Figure \ref{fig:react_distri} and value of interest lies in $[0.7, 2.4]$ seconds. 

%Although the independence assumption does not hold in real life cases, the computation of the probability of the accident occurring is independent of the calculation of severity (expected loss) of the accident. We refer readers to \cite{kodaka2003rear,carparameters} for more rigorous statistic analysis of the probabilities.
%and the formal one is not our concentration in this paper.

We analyze the risk by dividing  $[40,50]$ in to $10$ consecutive small intervals $[d^i_{l},d^i_{u}],i=1,\dots,10$, and $[0.7, 2.4]$ into $17$ consecutive intervals $[r^i_{l},r^i_{u}],j=1,\dots,17$. For each region consists of small intervals of $d$ and $r$, we use {\toolname} to verify safety or compute the worst case collision velocity. To compute the probability of the accident occurring, we need to compute the probability that each parameter lies in the given range.
For distributions shown in Figures~\ref{fig:dist_distri} and \ref{fig:react_distri}, the probability of the region $d \in [d^i_{l},d^i_{u}], r \in [r^j_{l},r^j_{u}]$ is $Pr(d \in [d^i_{l},d^i_{u}]) \times Pr(r \in [r^j_{l},r^j_{u}])$ if we assume the events $ d \in [d^i_{l},d^i_{u}]) $ and $r \in [r^j_{l},r^j_{u}]$ are independent.
For example, $Pr(41 \leq d \leq 42, 1.0  \leq r \leq 1.1) = 0.19 \times 0.139 = 0.026$. 

With the given braking profile and initial velocity of both cars, we can compute the worst case relative velocity for region of $d$ and $r$. We report the results in Figure \ref{fig:wcrvc}. The numbers correspond to each rectangle in the figure are the worst case relative velocities. For example, for the case $d \in [40,41]$ and $r \in [2.3,2.4]$, the worst case relative velocity $v_c$ is $17.5(m/s)$. We also plot the heat map of risks as the background of  Figure \ref{fig:wcrvc}, where the green rectangles with number $0$ correspond to the safe cases. Combined with the probability of occurrence, we can compute the expected velocity in the collision for Figure \ref{fig:distributions} to be 
$E[v_c] = \sum_{i=1}^{10}\sum_{j=1}^{17}Pr(d \in [d^i_{l},d^i_{u}]) \times Pr(r \in [r^j_{l},r^j_{u}]) \times v_c(i,j)  = 2.86 (m/s).$ Therefore, for AEB system with given braking profile and initial veclocity for each car, and given distributions for initial separations and reaction times, we can compute the risk defined in ASIL as the expected worst case relative velocity for the collisions.  

%The safety verification using {\toolname} for a single case takes $3$ to $50$ seconds on a Laptop with Intel Core i7-6000U CPU and 16 GB RAM. Together, it takes $5$ to $30$ minutes to generate a risk heat map.
  
% Single car running time?
% Runtime verification?

%%!TEX root = main.tex
\section{Integrated safety analysis}
 \label{sec:int}
\begin{figure*}[tbhp!]
    \centering
    \begin{subfigure}[b]{0.32\textwidth}
        \includegraphics[trim={0 0 4.5cm 0},clip,width=\textwidth]{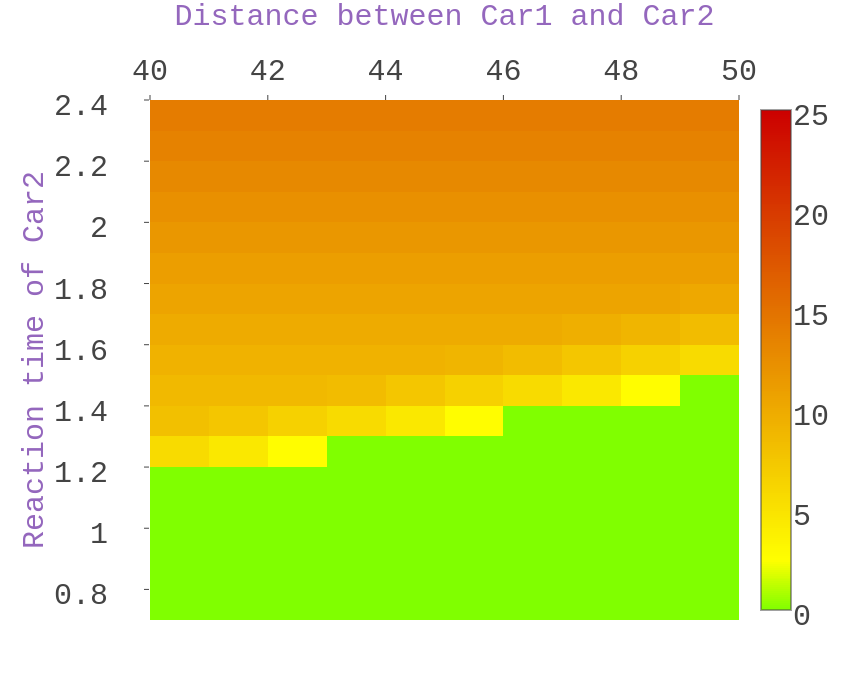}
        \vspace*{-10mm}
        \caption{\makecell{$\car 1$ and $\car 2$: mild brake, \\  initial velocity: $30(m/s)$}}
        \label{fig:tc1}
    \end{subfigure}
  %add desired spacing between images, e. g. ~, \quad, \qquad, \hfill etc. 
      %(or a blank line to force the subfigure onto a new line)
%    \begin{subfigure}[b]{0.32\textwidth}
%        \includegraphics[width=\textwidth]{figs/a1=a2=-10;v1=v2=30.png}
%        \caption{$\car 1$ and $\car 2$: medium brake}
%        \label{fig:tiger}
%    \end{subfigure}  
%    \vspace{5pt}
     %add desired spacing between images, e. g. ~, \quad, \qquad, \hfill etc. 
    %(or a blank line to force the subfigure onto a new line)
    \begin{subfigure}[b]{0.2635\textwidth}
        \includegraphics[trim={4.5cm 0 4.5cm 0},clip,width=\textwidth]{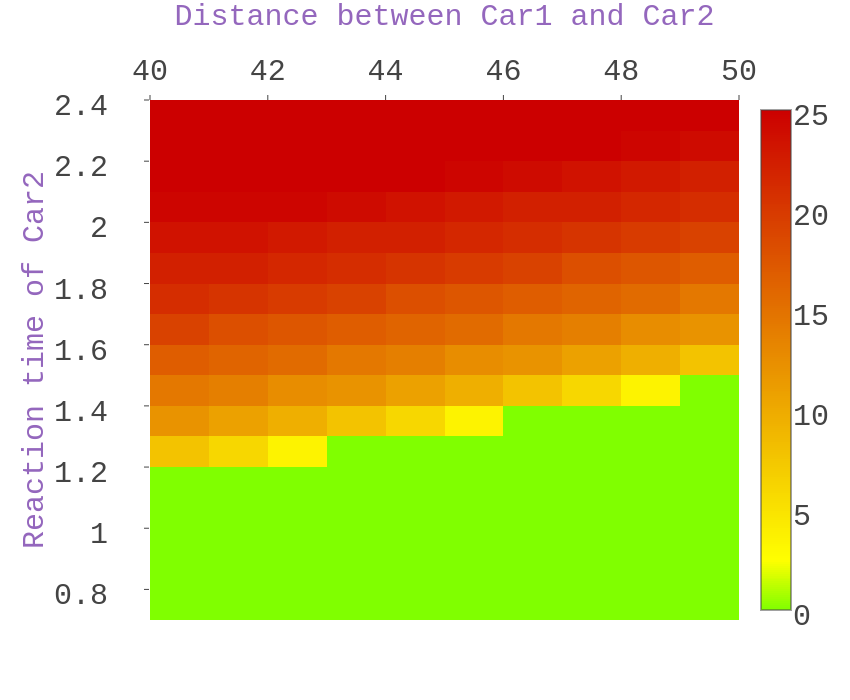}
	\vspace*{-10mm}
        \caption{\makecell{$\car 1$ and $\car 2$: hard brake, \\ initial velocity: $30(m/s)$}}
        \label{fig:tc2}
    \end{subfigure} 
     \begin{subfigure}[b]{0.32\textwidth}
        \includegraphics[trim={4.5cm 0 0 0},clip,width=\textwidth]{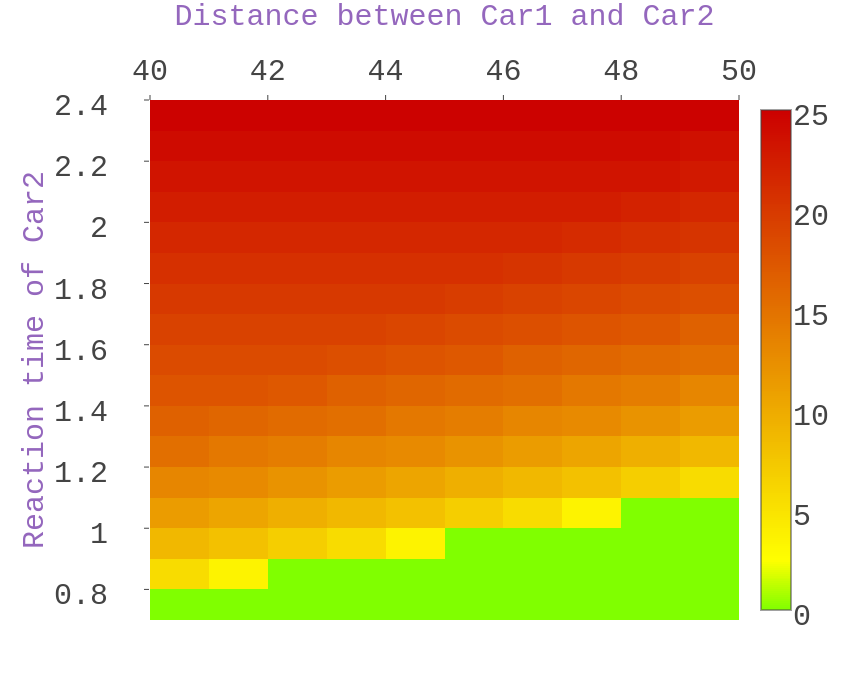}
        \vspace*{-10mm}
        \caption{\makecell{$\car 1$: medium brake, $\car 2$: mild brake, \\ initial velocity: $30(m/s)$}}
        \label{fig:tc3}
    \end{subfigure} 
%     \begin{subfigure}[b]{0.32\textwidth}
%        \includegraphics[width=\textwidth]{figs/a1=-8;a=-10;v1=v2=30.png}
%        \caption{$\car 1$: mild brake, $\car 2$: medium brake}
%        \label{fig:mouse}
%    \end{subfigure} 
%    \caption{Safety of two cars following scenario: varying declaration. In each subplot, fix the initial velocity to be $v_{1}(0) = v_{2}(0) = 30m/s$. }\label{fig:animals}
%\end{figure*}

\vspace{5pt}
%\begin{figure*}[tbhp!]
%    \centering
    \begin{subfigure}[b]{0.32\textwidth}
        \includegraphics[trim={0 0 4.5cm 0},clip,width=\textwidth]{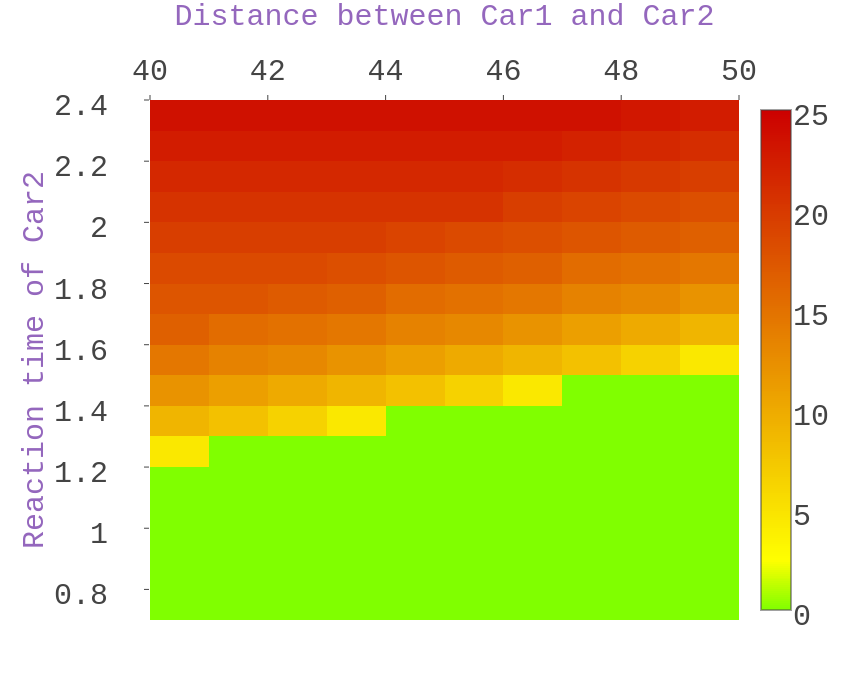}
        \vspace*{-10mm}
        \caption{\makecell{$\car 1$ and $\car 2$: medium brake, \\  initial velocity: $30(m/s)$}}
        \label{fig:tc4}
    \end{subfigure}
  %add desired spacing between images, e. g. ~, \quad, \qquad, \hfill etc. 
      %(or a blank line to force the subfigure onto a new line)
    \begin{subfigure}[b]{0.2635\textwidth}
        \includegraphics[trim={4.5cm 0 4.5cm 0},clip,width=\textwidth]{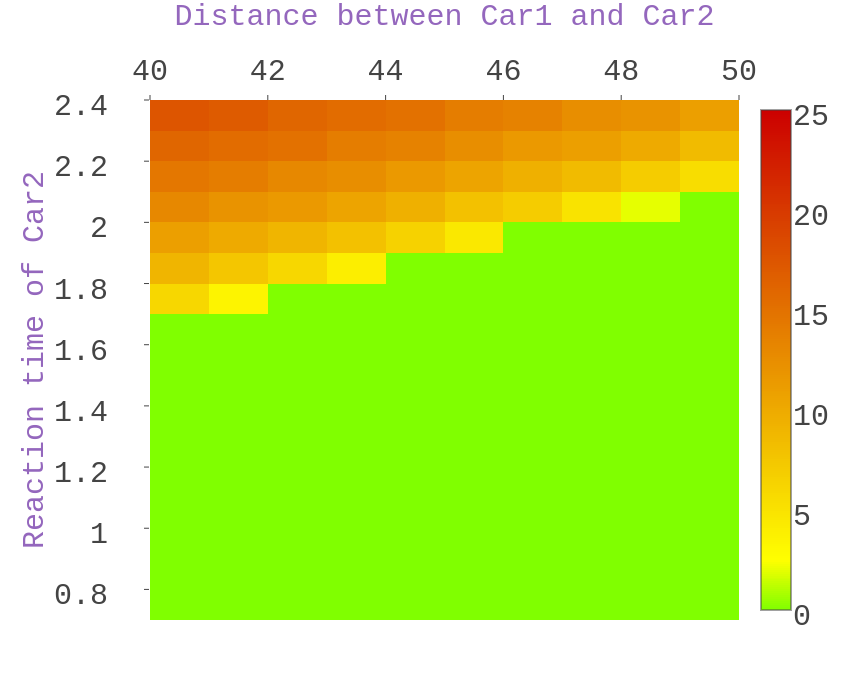}
        \vspace*{-10mm}
        \caption{\makecell{$\car 1$ and $\car 2$: medium brake, \\  initial velocity: $22(m/s)$}}
        \label{fig:tc5}
    \end{subfigure}  
%    \vspace{5pt}
     %add desired spacing between images, e. g. ~, \quad, \qquad, \hfill etc. 
    %(or a blank line to force the subfigure onto a new line)
    \begin{subfigure}[b]{0.32\textwidth}
        \includegraphics[trim={4.5cm 0 0 0},clip,width=\textwidth]{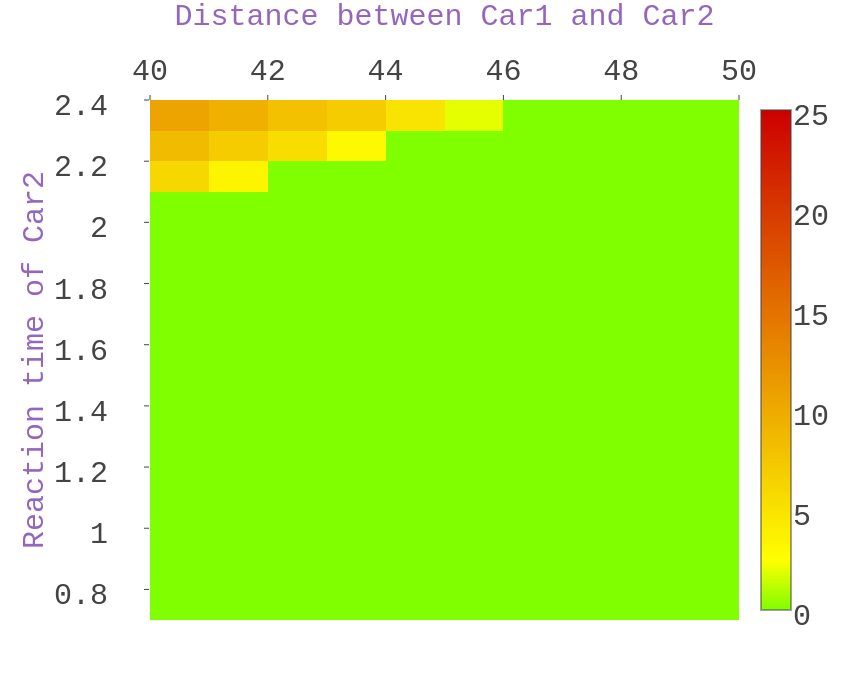}
        \vspace*{-10mm}
        \caption{\makecell{$\car 1$ and $\car 2$: medium brake, \\  initial velocity: $18(m/s)$}}
        \label{fig:tc6}
    \end{subfigure} 
%     \begin{subfigure}[b]{0.33\textwidth}
%        \includegraphics[width=\textwidth]{figs/a1=a2=-10;v1=v2=15.png}
%        \caption{Initial velocity: $v_1(0) = v_2(0) = 15m/s$}
%        \label{fig:mouse}
%    \end{subfigure} 

\vspace*{5pt}
      \begin{subfigure}[b]{0.33\textwidth}
        \includegraphics[trim={0 0 4.5cm 0},clip,width=\textwidth]{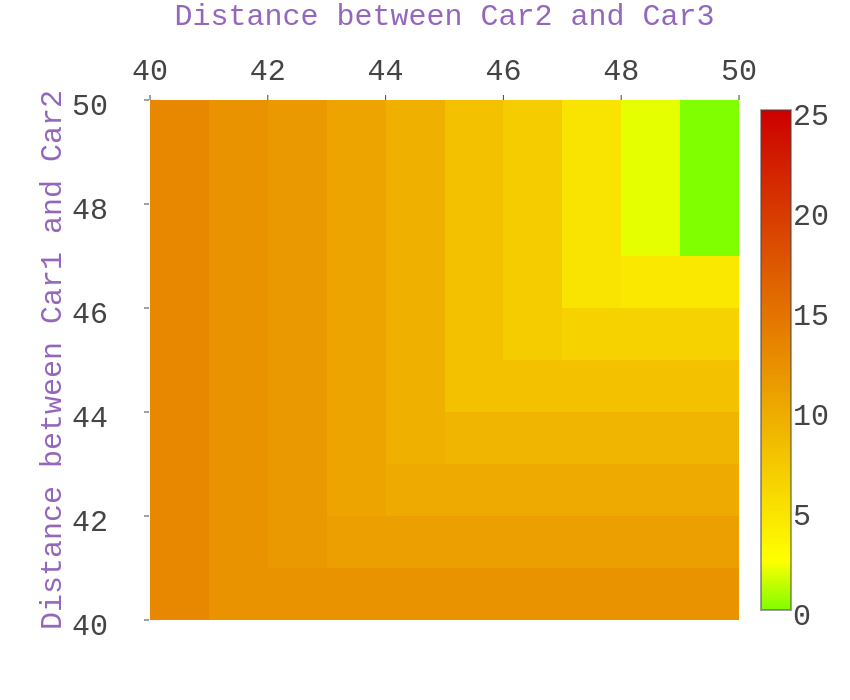}
\vspace*{-10mm}
    \caption{\makecell{\small Initial separation: $d_{23}$ vs $d_{12}$. \\ Fix the reaction time of both $\car 2$ and $\car 3$: \\ $r_2,r_3 \in [1.8,1.9](s)$.} }\label{fig:three cars dist}
\end{subfigure}
~~~~
 \begin{subfigure}[b]{0.39\textwidth}
    \centering
        \includegraphics[width=\textwidth]{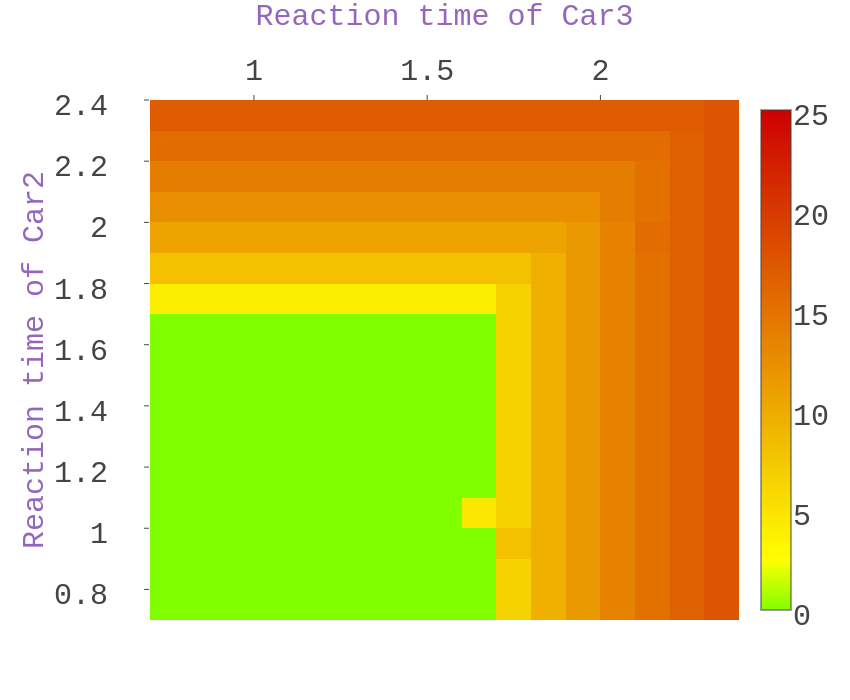}
        \vspace*{-10mm}
    \caption{\makecell{Reaction time: $r_{3}$ vs $r_{2}$.  \\ Fix the initial separation of both pairs of cars: \\ $d_{12},d_{23} \in [44,45](m)$}}\label{fig:three cars react}
\end{subfigure}

    \caption{\small Top row: two cars with different braking profiles and fixed initial velocities. Middle row left to right: two cars with decreasing velocities and fixed braking profiles. Bottom row: three cars with each car's deceleration is medium hard, and initial velocity is $22(m/s)$.}\label{fig:two_cars_hm}
\end{figure*}

Data-driven verification can be used to gain detailed insights about the safety of autonomous and ADAS systems under different scenarios and parameter variations. For the emergency braking system with two and three vehicles,  we have analyzed  hundreds of experiments; the summary of the  worst-case relative collision velocities computed from these experiments are shown in Figures~\ref{fig:two_cars_hm}.

We consider $3$ different braking profiles for each vehicle: mild, medium and hard. The average deceleration rate increases from mild to hard. The risk analysis can also be applied to any braking profile like the threshold braking and cadence braking used in Anti-lock Braking Systems. 
Figures~\ref{fig:tc1} and~\ref{fig:tc4} show the collision heat maps with fixed initial velocities but changing braking profiles for two  cars. From Figures~\ref{fig:tc1},~\ref{fig:tc2} and~\ref{fig:tc4}, we observe that if the lead  and the following cars have the similar level of braking, the safe regions are nearly invariant. However, with the increasing of the braking level, the severity (relative velocities) of collisions also increase. 
Comparing with Figure~\ref{fig:tc4}, we can see that if the lead car brakes harder than the follower, then as expected, the safety regions shrink rapidly. Moreover, the collisions are more severe than those in the previous case with both  cars braking equally hard. If the lead car brakes more gently (Figure~\ref{fig:wcrvc}), then the severity reduces quickly.
Therefore, qualitatively, it is safer for the following car to choose a braking profile harder or equal to the braking profile of the lead car. 

Figures~\ref{fig:tc4}-\ref{fig:tc6} show a sequence of collision heat map with fixed braking but changing initial velocities. As expected, both the area of the unsafe regions and severity of collisions decrease with reduction of the initial  velocities. The  analysis enables us to prove that, for example, the system is safe when the initial velocities of both cars are less than $17(m/s)$ for the given braking profiles and reaction times.

%\begin{figure}[tbhp!]
%    \centering
%      \begin{subfigure}[b]{0.7\textwidth}
%        \includegraphics[width=\textwidth]{figs/a1=a2=a3=-10;v1=v2=v3=22;rt=[18,19].png}
%    \caption{\makecell{Initial separation: $d_{12}$ vs $d_{23}$. \\ Fix the reaction time of both $\car 2$ and $\car 3$: \\ $r_2,r_3 \in [1.8,1.9](s)$.} }\label{fig:three cars dist}
%\end{subfigure}
% \begin{subfigure}[b]{0.7\textwidth}
%    \centering
%        \includegraphics[width=\textwidth]{figs/a1=a2=a3=-10;v1=v2=v3=22;dist=[44,45].png}
%    \caption{\makecell{Reaction time: $r_{2}$ vs $r_{3}$.  \\ Fix the initial separation of both pairs of cars: \\ $d_{12},d_{23} \in [44,45](m)$}}\label{fig:three cars react}
%\end{subfigure}
%\caption{Safety of AEB with three cars. In each subplot, each car's deceleration is medium hard, and initial velocity is $22(m/s)$.} \label{fig:three_cars_hm}
%\end{figure}

For the system with three cars, we consider scenarios with $4$ parameters: the initial separations $d_{12},d_{23}$ and reaction times $r_{2},r_{3}$. For visualizing the risk, we fix the range of $2$ parameters while varying the others. 
Fixing the reaction times of both $\car 2$ and $\car 3$ to be within the range $[1.8,1.9] (s)$, we analyze the change of safety envelope with respect to the change of $d_{12}$ and $d_{23}$. Figure \ref{fig:three cars dist} shows that the system is collision-free only when both the distances $d_{12}$ and $d_{23}$ are large enough. Compare Figure \ref{fig:three cars dist} with Figure \ref{fig:tc5} when all the cars have the same initial velocities and braking profiles. We can see the when the reaction time is between $[1.8,1.9](s)$, the safe distance change from $d>44 (m)$ for system with two cars to $d_{12}>47 (m), d_{23}>49 (m)$ for system with three cars. Therefore, with the increase of number of cars in a chain, the ``safe'' distance between any pair of cars increases as well.

Next, fixing the distance $d_{12},d_{23}$ to be within the range $[44,45](m)$, we analyze the change of safety envelope with respect to the change of reaction time $r_{2},r_{3}$. Figure \ref{fig:three cars react} shows that the cars are collision free only if both $\car 2$ and $\car 3$'s reaction time are short enough. Compared with Figure \ref{fig:tc5} again, when the distance between the cars are between $[44,45](m)$, the safe reaction time change from $r<1.9(s)$ for the two cars scenario to $r_{2}<1.7(s),r_{3} < 1.6(s)$ for three cars scenario. Both Figure \ref{fig:three cars dist} and \ref{fig:three cars react} show quantitatively that the safety envelope shrinks with the increase of number of cars in the system. 

As the running time for each scenario is $3-5$ seconds on a standard laptop, it takes $5-30$ minutes to generate a heat map, which suggest that similar analysis could be applied to more complicated scenarios with larger number of modes,  parameters, and more sophisticated ADAS systems.
%%!TEX root = main.tex
\section{Conclusion}
\label{sec:conclude}

We sketched recent developments in verification tools of CPS that combine formal reasoning with simulation data to effectively prove safety or estimate worst case accidents for automotive control systems.
%
%Formal verification can be used in autonomous driving control systems to guarantee safety or estimate worst case severity of the accident within short amount of time. In this paper, we presented a case study of straight road, car-following automatic emergency braking systems consisting of a sequence of 2 or 3 cars. We demonstrated how to use data-driven verification technology to provide a rigorous foundation for certification and risk analysis. 
%
Our case study with emergency braking  show that designers can use the tool for analyzing  autonomous driving and ADAS features under a variety of  traffic scenarios. Engineering the tools  to scale to bigger scenarios with many more modes and vehicles while achieving near  real-time performance  will be an natural and important  next step. 
%There are, of course, a lot of open problems with be solved to improve the performance of the tool. For example, the scalability of the verification technique with large number of variables, the generalization of the the approach to handle unbounded time properties, etc. 

%We use the distributions of the distance between vehicles on the straight road and the reaction time for the following cars as prior knowledge to analysis the safety and risk of different braking profiles. We chop up the large interval of separation distances and reaction time into small intervals, and perform data-driven verification for each case. The final result will be risk heat map with the values on the map to be the worst case relative velocity between the vehicles if there is a collision. By changing the initial conditions, we can conduct comprehensive analysis on the control systems.

%The data-driven verification technology presented in this paper is poised to take on the challenge to
%\input{related}
\section*{Acknowledgement}
We thank Joe Miller for pointing us to the emergency braking  case study.
This case study would be impossible without our collaboration with Mahesh Viswanathan
on {\toolname}.

\bibliographystyle{ieeetr}
\bibliography{citation}

%\newpage
%\appendix
%\input{app_basic}
%\input{appendix_new}

\begin{IEEEbiography}
{Chuchu Fan} is a PhD candidate at Electrical and Computer Engineering Department, University of Illinois at Urbana-Champaign. She received her Bachelor degree from Automation Control Department, Tsinghua University in 2013. Her research interests are in verification of nonlinear hybrid systems.
\end{IEEEbiography}

\begin{IEEEbiography}
{ Bolun Qi} is a master student at Computer Science Department, University of Illinois at Urbana-Champaign, where he received his Bachelor degree as well in 2016. His research interests are in development of formal verification tools.
\end{IEEEbiography}

\begin{IEEEbiography}
{Sayan Mitra} is an Associate Professor of Electrical and Computer Engineering at the University of Illinois at Urbana-Champaign. His research interests are in formal methods, distributed systems and hybrid control systems with applications in automotive, medical, and robotic systems. He received a PhD from  MIT and held a post-doctoral fellowship at California Institute of Technology. He has held visiting positions at Oxford University and the Air Force Research Laboratory at New Mexico. He received the National Science Foundation's CAREER Award, Air Force Office of Scientific Research Young Investigator  Award, and the IEEE-HKN C. Holmes MacDonald Outstanding Teaching Award.
\end{IEEEbiography}

\end{document}